\documentclass[9pt,twocolumn,twoside]{osajnl}
\usepackage{float}
\journal{jocn} 

\setboolean{shortarticle}{false}
\title{Network Aware Compute and Memory Allocation in Optically Composable Data Centres with Deep Reinforcement Learning and Graph Neural Networks}

\author[1,*]{Zacharaya Shabka}
\author[1]{Georgios Zervas}

\affil[1]{Optical Networks Group, Department of Electronic and Electrical Engineering, University College London, Roberts Building, Torrington Place, London, WC1E 7JE, United Kingdom}

\affil[*]{Corresponding author: zacharaya.shabka.18@ucl.ac.uk}

\begin{abstract}
Resource-disaggregated data centre architectures promise a means of pooling resources remotely within data centres, allowing for both more flexibility and resource efficiency underlying the increasingly important infrastructure-as-a-service business. This can be accomplished by means of using an optically circuit switched backbone in the data centre network (DCN); providing the required bandwidth and latency guarantees to ensure reliable performance when applications are run across non-local resource pools. However, resource allocation in this scenario requires both server-level \emph{and} network-level resource to be co-allocated to requests. The online nature and underlying combinatorial complexity of this problem, alongside the typical scale of DCN topologies, makes exact solutions impossible and heuristic based solutions sub-optimal or non-intuitive to design. We demonstrate that \emph{deep reinforcement learning}, where the policy is modelled by a \emph{graph neural network} can be used to learn effective \emph{network-aware} and \emph{topologically-scalable} allocation policies end-to-end. Compared to state-of-the-art heuristics for network-aware resource allocation, the method achieves up to 20\% higher acceptance ratio; can achieve the same acceptance ratio as the best performing heuristic with $3\times$ less networking resources available and can maintain all-around performance when directly applied (with no further training) to DCN topologies with $10^2\times$ more servers than the topologies seen during training. 
\end{abstract}

\setboolean{displaycopyright}{true}

\begin{document}

\maketitle

\section{Introduction}
\label{section:intro}
Contemporary data centre network (DCN) architectures are based on (opto-)electronically packet-switched (EPS) networks. In a typical Cloud computing model, large tasks requiring more than one server's worth of resources for a long period of time can be distributed across numerous servers, which are all connected via this underlying EPS infrastructure. These allocations can not be done across single inter-server resource pools, instead requiring the task to be split into smaller tasks where each will be allocated resources from and run on a single server \cite{Hadary2020}.

Firstly, bandwidth is limited by the bandwidth per-port of the opto-electronic switches which is fixed per-model. Popular EPS switches typically support a per-port bandwidth at the order of $\mathcal{O}(1Gb/s)-\mathcal{O}(10Gb/s)$, whereas intra-server communications (e.g. L1-cache access by the CPU) often operate at the order of $\mathcal{O}(Tb/s)$. For resources to access each other remotely, a large number of ports would be required for just a single pair of devices and networking component costs would increase. Such devices also have fixed bandwidth, meaning higher bandwidth servers require network infrastructure replacement or must be run sub-optimally.

Secondly, the unpredictable queuing patterns in packet-switched networks lead to non-deterministic latency. Compute mediums (i.e. CPU, GPU, RAM) co-located on the same server exchange information at very high rates. For example, L1 cache latency on high-end desktop CPUs exist in the $\mathcal{O}(ns)-\mathcal{O}(10ns)$ range. Application performance is strongly dependent on compute-memory latency \cite{popescu2019}. EPS networks are incapable of consistently supporting standard application performance given typical forwarding latency is non-deterministic and in the range $\mathcal{O}(10\mu s)-\mathcal{O}(100\mu s)$ . These features of EPS DCNs lead to two primary resource allocation limitations when allocating large quantities of resources to single tasks for some extended period of time:

\textbf{Resource fragmentation} means that resources can become `stranded' on a particular server, neither in use by any applications running on that server, nor accessible by those running on any others; while sufficient resources for some task may exist collectively across several servers in a DC, they cannot be co-assigned to said task and are effectively wasted. 

\textbf{Inflexible resource-pooling} relates to the strict upper limit in resource pool size. In modern EPS-based DCs, this is limited to the amount of resources that can be hosted on a single server, since inter-server resources cannot communicate with the means required to not disrupt application performance. Using pools larger than this requires further consideration about how applications can be effectively divided into sub-applications and distributed, often incurring some runtime-overhead.

Resource disaggregated DCs are a proposed DC architecture supporting a network architecture that provides sufficient bandwidth and latency guarantees for resource pools to be defined across servers on which long-lived resource-hungry applications can be run with local-like performance. This would reduce fragmentation, as well as increase pooling flexibility. Such architectures can in fact be built using off-the-shelf commodity hardware such as commercially available optical-circuit switches \cite{polatis_switch, Zervas17, Zervas18,Mishra21}. However, since both server- and network- resources need to be explicitly provisioned in order to allocate both compute and connectivity, allocation is more complex as decisions need to be made across the product of both of these domains, rather than only server resources as in conventional resource management frameworks \cite{FacebookTupperware, Isard2009, Verma2015, Schwarzkopf2013}. This paper will refer to this requirement as \textit{network aware resource allocation}.

This paper shows that deep reinforcement learning (DRL) with graph neural network (GNN) based policies can learn very effective network aware allocation policies end-to-end. Acceptance rate, CPU utilisation and memory utilisation are improved by up to 19\%, 24\% and 22\% respectively compared to state-of-the-art heuristics. Furthermore the DRL-based method achieves approximately the same performance as the best heuristic achieves when it is using $3\times$ more network resources. While the method is trained on small DCN topologies with $\mathcal{O}(10^1)$ servers, the GNN-based policy architecture is topology-size agnostic. Because of this it can be directly applied to topologies with $10^2\times$ more servers than seen during training and maintain it's allocation performance without further training required. Following this, a discussion on interpreting the learnt policy is presented, indicating that the method is flexible under changing network resource profiles and generally more adaptable than the heuristics. 

%Code used for this work is available \underline{\href{https://github.com/zawaki/allocation_neurips_2021_mlfs}{here}}.
\section{Previous Work}

Deep learning has been leveraged to tackle numerous online combinatorial optimisation (CO) problems. Pointer networks were introduced as a supervised means of doing so in \cite{Vinyals15}, where this premise was integrated into a DRL framework in \cite{Bello16}. Similarly, a similar methodology was applied to a simple multi-resource cluster allocation problem in \cite{Mao2016}. Following this, \cite{Dai2018} proposed that when solving combinatorial optimisation problems defined on graphs (e.g. travelling salesperson, max-cut and so on) it should be useful to account for some topological information in the optimisation process. A DRL-based framework for solving CO problems where policies were modelled with GNNs was proposed where GNNs are used to generate node-embeddings which can be used in some selection process to more accurately construct a solution to the underlying problem. Further iterations of this framework have shown it able to scale to graphs with the order of $\mathcal{O}(10^7)-\mathcal{O}(10^9)$ nodes with similar orders of edges \cite{ying2018,Akash2019,Zhuwen2018}, perform competitively against exact solvers \cite{barrett19}. This architecture has been used to effetively solve a number of network and computer-system based optimisation problems, such as distributed machine learning \cite{addanki2019placeto}, cluster management with dependency-structured tasks \cite{Mao2019}, optical routing \cite{Almasan2019} and virtual network embedding \cite{Yao2018, Yan2020}.

Network aware algorithms for optically composable disaggregated data centres are described in \cite{Yuan18,Zervas17,Zervas18}. These works augment a breadth-first-search procedure to recursively discover sub-networks that can support the required compute, memory, storage, bandwidth and/or latency by some given request. They show advantages over traditional packing algorithms such as best fit, but suffer from poor scalability since they are exhaustive in the worst case. A bandwidth-aware multi-resource cluster allocation (and scheduling) method is described in \cite{Grandl14}, where servers are ranked based on server-local compute and network resources. However, this has limited exposure to the network as it does not consider network resources multiple hops away from the server. Previous work has shown that DRL with GNN-based policies can be used to learn network aware resource allocation algorithms in composable data centres which are both high performing and scalable \cite{shabka21}. This work continues the examination of these architectures in a similar experiemental setting, but analyse more extensively the nature of the policy that is learnt. Specifically, our analysis is extended to the network usage, fairness and general nature of the decision making by the DRL agent. In this way we seek to understand more intuitively what the agent is doing, as opposed to a more simple observation of improved long-term allocation outcomes. We discuss new results about how each tier's network resources are used by the agent (and comparative methods) across the test topologies, as well as analyse the relationship between request size and how the agent's allocations are distributed within racks, between racks and between clusters.

\section{Background}
\label{section:problem_spec}

\subsection{Deep Reinforcement Learning}

Reinforcement learning relates to the study of how to find optimal behavioural policies in dynamic environments. When rewards are returned to an actor based on the effect that their action had on the environments state, the goal of a reinforcement learning problem is to find a policy that maximises the reward achieved by an agent over time. A environment is formally described by a \textit{Markov Decision Process} (MDP) defined as a tuple $<S,A,R_a,T_a>$ where: $S$ is the set of all possible states that the MDP can be in, $A$ is the set of all possible actions that some actor can take in this environment, $R_a$ is a function describing the reward yielded when an agent is in state $s$, takes action $a$ and ends up in state $s'$, and $T_a$ is a function describing the probability of an agent being in state $s$, taking action $a$ and ending up in state $s'$. Episodes can also be episodic meaning that there is some state after which the environment no longer changes state (e.g. a check-mate position in a chess game environment). A policy is described as some function, $\pi(s) \rightarrow a$ which maps states to actions, and an optimal policy is one which - if followed - will yield the highest possible reward in that MDP. A very extensive description of RL from first principles can be found in \cite{sutton18}.

Deep reinforcement learning is an extension of this generic problem description, to the case where $\pi(s)$ modelled using deep neural networks. In contrast to older methods (such as table-based policies) has allowed for much greater complexity to be learnt by the policy, as well as greater generalisability when a policy is exposed to a previously unseen state. Such developments have seen DRL exceed human performance in considerably complex tasks like the board game of Go and the video game Starcraft \cite{silver17,Vinyals2019}.

\subsubsection{Graph Neural Networks}

Graph neural networks extend standard neural network (NN) architectures to graph-structured data, where data points are nodes and any relationships (e.g. a friendship between two people in a social network) are represented by edges. GNNs attempt to account for topological information as well as the raw data during learning tasks. This is accomplished by means of a \textit{message passing procedure} where node (and possible edge) information is propagated through the network via nearest-neighbour exchanges and aggregated at nodes. These aggregations are then used as a new representation of each node, which can be processed by some NN structure. This procedure can be repeated (for all nodes simultaneously) in order to propagate information further through the network, and as such the final output of the GNN for each node is one which accounts for information about itself, it's neighbours and local network region. This procedure can be represented by the equation:
\begin{equation}
    h_v = g(v,\sum_{v^{'} in N(v)}f(v^{'},e_{v,v^{'}}))
\end{equation}
where $g$ and $f$ are (learnable) functions, $v$ is the information at node $v$, $e_{v,v^{'}}$ is the information at the edge connecting nodes $v$ and $v^{'}$, $N(v)$ is the set of all one-hop neighbours of node $v$ and $h_v$ is the new representation of node $v$ after a single message passing procedure has been applied. In the case of a GNN, $g$ and $f$ will generally be implemented with a neural network. GNN architectures have shown to yield richer node embeddings than classical node-embedding methods (e.g. PageRank) and can outperform these methods in statistical tasks such as graph clustering or node classification \cite{Velickovic2018, Hamilton2017, Kipf2017}.

\subsubsection{Combinatorial Optimisation}

% The optimality of a solution is typically determined by some objective function which maps solutions to numerical values, such that finding a solution that maximises or minimises this function is equivalent to finding an optimal set. 

Combinatorial optimisation (CO) problems describe scenarios with a finite set of items, where some optimal sub and/or ordered set of items (often termed the \textit{solution}) must be determined. Typically there will be some validity criteria for a solution such that some solutions do not have a value (e.g. a solution that does not traverse every node once and only once in a travelling salesperson problem is not valid). Formally, a CO problem can be described by a set of instances, $I$ (an instance could be a particular set of bins in a bin packing problem); for some instance $x \in I$, $f(x)$ is the set of valid solutions and $m(x,y)$ (termed the objective function) maps some valid solution $y \in f(x)$ to some number; the goal of a CO problem is to find, for some $x \in I$, a solution $y^{'} \in f(x)$ such that $m(x,y^{'})$ is either minimised or maximised (depending on the problem).

There are no formal constraints on how solutions can be constructed. They can be either determined entirely and then evaluated (as in some exact solvers) or they can be iteratively constructed by adding items one at a time and continuously evaluating whether the current solution is valid, and if so what is it's objective value. For graph-based CO problems (e.g. routing, min-cut etc), solutions can often be built by iteratively adding nodes from the graph to the solution set (as will be implemented in this paper).

\section{Problem}
In this paper we present the problem as online network aware resource allocation in dissaggregated data centre systems as a MDP and show that GNN-based DRL can be leveraged effectively to solve this problem. This section will describe how the MDP is defined including the request/resource allocation dynamics of the underlying DCN, the learning architecture used to model the policy, the experiments carried out and the baselines used for comparison.

\subsection{Defining the Markov Decision Process}
\label{section:mdp}
\textbf{Environment:} The environment consists of a set of servers, a network interconnecting these servers via some network switches and requests which arrive one by one. This DCN environment is also visualised in \ref{appdx:topology}.

The DC/DCN consists of the servers + network and is represented by a graph, $G(V,E)$. Each server (represented by the set of nodes $V \in G$) has an associated resource vector. Specifically, in this problem the CPU and memory resources are accounted for so that $[v_{cpu},v_{mem}] \forall v \in V$, where $v_{cpu}$ \& $v_{mem}$ represent the available CPU and memory resources of server $v$ at any given moment.

Similarly, each switch has a particular number of input and output ports. This is represented by proxy using edge features, such that each edge has a number of distinct channels, also denoted by a resource vector $[e_{ch}] \forall e \in E$ where $e_{ch}$ is the number of available channels in that link. This represents a scenario where a certain number of ports on a switch are reserved for a particular server who's direct link to that switch has a certain number of channels.

\begin{table}[h!]
\centering
\caption{Oversubscription and number of channels per link for each topology. `Bottom-top oversubscription' refers to the oversubscription from the servers to the top tier of switches (tier-3). `Oversub' refers to the oversubscription at the interface between that tier and the tier below it (hence Tier-1 does not have an `oversub' value. In this work we used topologies of this structure with $n \in \{8, 16, 32\}$.}
\label{table:oversub}
\begin{tabular}{lllllll}
\multicolumn{7}{c}{\textbf{Oversubscription, channels-per-link for network tiers}} \\
\toprule
\multicolumn{7}{c}{\textbf{Bottom-top Oversubscription}} \\
% \multicolumn{9}{c}{\small{\textbf{$n = channels\ per\ tier\mhyphen1\ link$}}} \\
\midrule
{} & \multicolumn{2}{c}{1:16} &          \multicolumn{2}{c}{1:8} &          \multicolumn{2}{c}{1:4} \\
{} & \multicolumn{2}{c}{Ovsub, Chan} & \multicolumn{2}{c}{Ovsub, Chan} & \multicolumn{2}{c}{Ovsub, Chan}  \\
\midrule
Tier-1 & \multicolumn{2}{c}{-, $n$} & \multicolumn{2}{c}{-, $n$} & \multicolumn{2}{c}{-, $n$}  \\
Tier-2 & \multicolumn{2}{c}{1:4, $2n$} & \multicolumn{2}{c}{1:2, $4n$} & \multicolumn{2}{c}{1:2, $4n$}  \\
Tier-3 & \multicolumn{2}{c}{1:4, $\frac{1}{2}n$} & \multicolumn{2}{c}{1:4, $n$} & \multicolumn{2}{c}{1:2, $2n$}  \\
\bottomrule
\end{tabular}
\end{table}

Requests arrive at the RDDC one at a time and are not seen in advance, where $R = [r_{cpu},r_{mem},r_{t}]$ represents the CPU, memory and holding time requirements for that request. A valid solution consists of a set of nodes, $V^{'}$, must be found such that $\sum v_{x} \in V^{'} \geq r_{x}, x \in \{cpu,mem\}$ and a set of $\frac{|V^{'}|(|V^{'}|-1)}{2}$ distinct paths can be found to guarantee all-to-all connectivity between all $v \in V^{'}$. A good solution to this problem is one which maximises successful allocations over time.
% \begin{enumerate}
%     \item $\sum v_{x} \in V^{'} \geq r_{x}, x \in \{cpu,mem\}$;
%     \item a set of $\frac{|V^{'}|(|V^{'}|-1)}{2}$ distinct paths can be found to guarantee all-to-all connectivity between all $v \in V^{'}$.
% \end{enumerate}
% A solution to this problem is considered to be one which can maximise the number of allocations made over time, given a series of $N$ requests.

%copy-paste start
\textbf{Episode:} An episode is defined as receiving $N$ requests. For each request the agent will iteratively choose servers until either it has successfully allocated a request or it's solution is invalid. When a new request arrives, if there is enough raw resources available in the DCN to theoretically allocate it, the agent is prompted to attempt to do so. If a valid solution is generated (there are not enough network resources available to connect all servers in the allocation) then this request is dropped and a new request is fetched. We do not model queuing as this involves another algorithmic domain (to balance new and queued request with priority structures etc) and in this work we wish to focus on the network-awareness/congestion dynamics of the underlying system. Moreover, queuing/prioritising can be handled in parallel to resource allocation decisions so can be considered separately.

% The agent is prompted to try to allocate servers when a new request has been received \textit{and} if there are enough resources present in the DC environment as a whole for it to theoretically be allocated. As such, agents will never be punished for attempting to allocate requests which are in fact impossible to allocate, since this would create trivial and misleading reward signals. Instead, if a request asks for more than what is currently available the agent will wait, and each time an already allocated requests finishes and de-allocated it will check if there is not enough resources available and attempt to allocate again.

\textbf{State:} given an awaiting request $R$ = $\{r_{cpu},r_{mem},r_t\}$ (CPU, memory and holding-time requirements respectively) the MDP's state is 
\begin{equation}
    s = \{G(V_{norm},E),r_{t},U_{cpu},U_{mem}\}
\end{equation}
where, $V_{norm}$ = $\{\frac{v_{x}}{r_{x}} \forall v_{x} \in V\},x \in \{cpu,mem\}$
and $U_{x}$ is the RDDC-global utilisation of resource $x \in \{cpu,mem\}$. This combines both node-, edge- and graph- level resource information within the state representation.

Feature normalisation is important in machine learning problems to ensure that certain features with larger absolute values/variation do not disproportionately influence policy updates during training. Furthermore, normalised state representation ensures that policies should be robust under testing on similar environments differing only be absolute scales (e.g. a DCN with 4 CPU units-per-server vs one with 16 units-per-server, or a scenario where requests are between 1-8 servers vs one where they are between 8-16). As such, each server's resource values are normalised with respect to the current requested amount of that resource. Similarly, link-resources are normalised with respect to the maximum initial amount on any link in the DCN. This ensures that the policy is exposed to the DCN in a way that is feature-scale agnostic, as well as agnostic to the relative scale of request quantities and server-resource quantities.  

% \textbf{Action:} A server $v \in V$ is chosen. K-shortest paths between $v$ and all nodes previously allocated to that request are found; if constraint 2 in section \ref{section:problem} cannot be satisfied by any of these paths for a node-pair, the allocation fails and the request is dropped. If constraint 2 is satisfied, then $max(v_{x},r_{x})$ from $v$ ($x \in \{cpu,mem\}$) and one channel per-link from the first path with sufficient resources for each pair is allocated to that request, before another node is chosen. If constraint 1 and 2 are both satisfied then after the previous step the allocation is successful and a new request is obtained. 
\textbf{Action:} A server $v \in V$ is added to $V^{'}$. If constraint 2 defined previously cannot be satisfied (when k=3 shortest paths are tried per node-pair) the allocation fails. If constraint 2 is satisfied, then $max(v_{x},r_{x})$ - the maximum that can be provisioned from this server up to the limit of how much is needed and how much is present at that server - from $v$ ($x \in \{cpu,mem\}$) and one channel-per-link per-path per-node-pair are allocated. If constraint 1 \& 2 are satisfied the allocation is successful. Per request, actions are taken until an action succeeds or fails an allocation, at which point a new request is fetched (until termination). 

\textbf{Reward:} An agent will receive $\alpha$ if the action is successful, $\beta$ if the action is unsuccessful, $\gamma$ otherwise where $10$, $-10$ and $0$ were used for $\alpha$, $\beta$ and $\gamma$ respectively. Rewarding 0 for intermediate decisions (i.e. when servers have been chosen but the request has neither failed nor been accepted) ensures that the reward is agnostic to the allocation choices. More generally, the reward is kept very simple and only based on allocation success or lack thereof for each request. This is done to minimise the influence on the kind of policies that the agent might learn. For example, if small negative intermediate rewards were given to incentivse allocating across as few servers as possible, this would likely not allow the agent to explore policies where allocating across lots of servers is sometimes preferable. More generally, one of the main motivations behind this work is that hand-designing heuristics is non-intuitive for complex systems or problems. As such,by simplifying the reward structure as much as possible and relating it only to a generic high level goal (allocate as many requests as possible over time), the designers are not meaningfully influencing the specific policy that the agent learns and allow it a more arbitrary exploration of possible solutions during training.

%copy-paste end

%\subsection{Assumptions and simplifications in the environment}

%Queuing or any notion of prioritisation are not accounted for in this work. While these are indeed relevant features of a real DC system, they do not necessarily feature in the specific domain of resource management that is addressed in this paper - choosing servers given a resource requirement. Many methods, including those employing similar learning architectures to that used in this work, independently address the complexities of queuing and prioritisation of tasks in scheduling environments. Though important, these problems can be addressed alongside the server-allocation addressed here but do not need to be co-addressed. As such we focus soley on the network-aware server-selection process required in resource disaggregated DC systems.

%The harshest possible connectivity requirements - all-to-all - are imposed on all requests. In reality, not every request necessarily needs this degree of server-interconnectivity. However, we explore a scenario here where resource pools are allocated to users for arbitrary use. Maximal connectivity between all servers in the allocation allows for this, though may be uneccesary for certain types of processes with more limited network requirements (e.g. MapReduce). This has the benefit of both simplifying the problem as well as maximising the importance of network-awareness in associated allocation scenarios.

\subsection{Defining the deep reinforcement learning model}

%copy-paste start
The learning model consists of a GNN (based on the GraphSAGE architecture \cite{Hamilton2017}) and 2 deep neural networks (DNN) which we refer to as DNN\textsubscript{1} and DNN\textsubscript{2}. The GNN acts on $G(V_{norm},E)$ to generate embeddings of each node in the topology. DNN\textsubscript{1} outputs a high dimensional representation of $[r_{t},U_{cpu},U_{mem}]$. DNN\textsubscript{2} then calculates logits for each node in the RDDC graph based on an input of the concatenation of the GNNs embedding of that node, the output of DNN\textsubscript{1} and the element-wise mean of the embeddings of the nodes that have already been allocated to that request (or a zero-vector if the request has just been received and nothing has been allocated yet). These logits are passed through a softmax function to specify the probability of choosing each node. This model is used to approximate a policy, trained using the proximal policy optimisation (PPO) RL algorithm, and is implemented using RLlib and Deep Graph Library \cite{rllib,dgl}.

\begin{figure}[h!]
    \centering
    \includegraphics[width=0.5\textwidth]{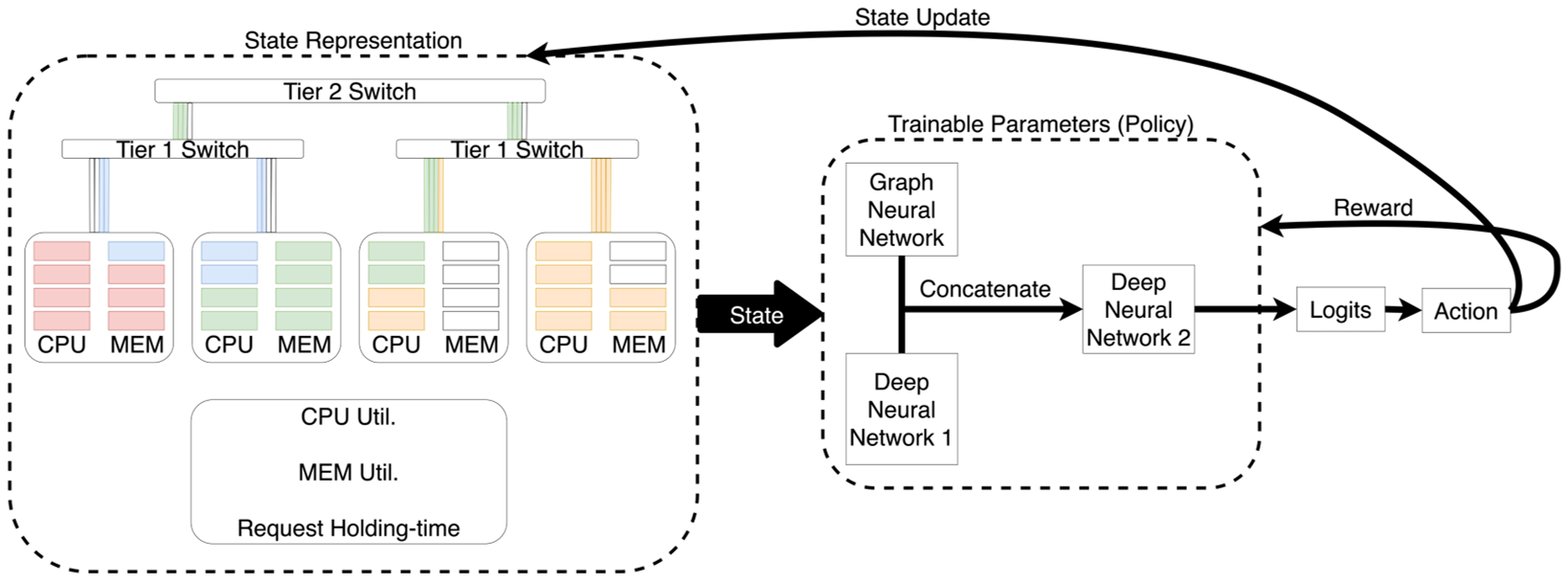}
    \caption{High-level diagram of the RDDC + model + RL feedback loop implemented in this work.}
    \label{fig:model}
\end{figure}

The GNN used a distinct mean-based aggregator for each layer, where messages exchanged during the message passing process are aggregated like:
\begin{equation}
    v = \frac{1}{|N(v)|+1} \sum_{x \in M_{v}} W_{i}(x)
\end{equation}
where $v$ is a node (embedding) in the RDDC graph $G(V,E)$, $N(v)$ is the set of the one-hop neighbours of $v$, $M_v$ is the set of messages received by $v$ from it's neighbours and $W_i$ is a neural network associated with the $i^{th}$ layer of the GNN.

The GNN outputs embeddings of each node in 16 dimensions, and 3 layers were used so that information from the top of the network can be accounted for in the embeddings of the servers.

We compare our model against 3 baselines (Tetris, NALB, NULB) from previous work as well as a random allocation policy \cite{Grandl14,Yuan18,Zervas18}.
%copy-paste end

\section{Experimental Setup}
\label{section:nara_experiment}

\subsection{Training and Testing}
%copy-paste start
Servers initially have 16 units of both CPU and memory resources. We use 12 different 3-tier as detailed in table \ref{table:oversub}. This is visualised in Fig. 7 in Appendix A.
% Servers are in the same rack if they are directly connected to the same set of tier-1 switches, and racks are in the same cluster if they are directly connected to the same set of tier-2 switches.
The values for channels-per-link in tier-1 (i.e. max number of other servers that a server connect to) is \{8,16,32\} and the set of bottom-to-top oversubscription ratios is \{1:4,1:8,1:16\}, where tier-2 and tier-3 channel values are set to ensure these ratios given the number of channels in a tier-1 link. Higher oversubscription (lower ratio) imposes a stronger mandate for rack-locality on allocations due to limited upper-tier network resources. Less tier-1 channels per server limits how many servers an allocation can be spread across, since all server within an allocation must be interconnected. Topologies used for training and testing have 64 nodes (2 clusters $\times$ 2 racks $\times$ 16 servers). Trained models were also tested on graphs the order of $10^2$ times larger with respect to number of nodes (8 clusters $\times$ 8 racks $\times$ 16 servers) with episodes of 2048 requests. Requests are uniformly sampled with a maximum request size of 8 full-servers worth of resources in each domain (128 units), and their holding times are sampled such that the average offered load on the RDDC system is 95\% of all CPU and memory resources. Separate agents are trained for each topology, and tested against each baseline on that same topology. 

Training episodes are terminated after 32 requests have been received (successfully or otherwise) by the agent, and testing episodes on the smaller topology are terminated after 128 requests have been received. shorter training episodes are generally more desirable since it is easier for the agent to receive meaningful reward signals \cite{Pohlen18,Parsonson22} (provided that the shortness of the episode is not so much so that some important dynamic/feature of the underlying environment is not experienced by the learning agent). On the other hand, in the scenario explored here, any trained agent deployed in some real DC scenario would be operating in an effectively infinitely long episode (i.e. it will continue to allocate resources to requests as long as the DC environment it manages is in use). As such it is important to ensure that the policy learnt is not limited in performance to just the short episodes seen during training. The training episode length is chosen long enough for all tested baselines to be at a `stable' performance level for the majority of the episode. For a short episode, much of the time is spent allocating the first requests when most of the resources are still available (sometimes referred to as `warm up'). While warm up strategy is important (bad decisions early on can cascade into long term inefficiencies), the performance statistics at the end of this phase are not indicative of long term performance where utilisation is consistently high as more requests are received. Testing episodes are therefore longer than training ones to emphasise performance on the long-term dynamics of the system (more akin to real DC system operation). Testing episodes are seeded in the same way for each baseline, so all methods are exposed to the same set of requests (per test) received in the same order. Each test is run 5 times and results presented (where a single value per topology is shown) are the average of these 5 runs. 

% This principle also applies to the tests on the much larger topologies with $\mathcal{O}(10^3)$ servers. Given the large amount of resources present in these DCs, it would take a very large number of requests to warm up the DC, and even more so to explore the longer-term dynamics. This means that such an environment is not desirable for training as noted above. After training on the smaller DC topology, the agent is tested directly on this larger topology with long episode lengths ($\mathcal{O}(10^3)$) requests. This allows for both the transferability (from small to large DC topologies) and the long-term allocation dynamics of the allocation policy to be explored. It should ideally be the case that, provided that the basic features of the DC are unchanged under scale-up (e.g. oversubscription), the policy would perform similarly on the larger DC topology as it does on the smaller one. 

As noted, tests are also implemented on scaled up ($\mathcal{O}(10^3)$ nodes) versions of each respective topology. Much longer episodes are required here since warm up takes much longer (since there is $\mathcal{O}(10^2)\times$ more resources in the DCN). Additionally, in order to maintain a high offered load the holding times are increased appropriate so that resource requirements build up in the system and allocation becomes harder as more requests arrive. More crucially, this test is done to explore the suitability of this method in a real DCN allocation scenario. Server clusters in large enterprise computer networks are of the order of $\mathcal{O}(10^3)$ servers and above. As such scalability to topologies of at least this scale is necessary. Furthermore, while a small test cluster may be feasibly reserved for experimentation \cite{Roy15}, full-scale experimentation is not possible as this would require halting all or much of the services provided by the cluster (since the experimental allocation techniques would be unsuitable for service level requirements). Where a sufficiently accurate simulation of DCN patterns is not available - which it often isn't \cite{parsonson21,Sharma2011,Cortez2017} - and a small cluster is reserved for experimentation, any algorithm developed on the small cluster must be consistent with respect to performance when transferred to the larger one.

All methods are evaluated on the basis of three metrics; acceptance ratio, CPU utilisation and memory utilisation. Acceptance ratio refers to the proportion of all requests received by some allocation method that were successfully allocated. CPU and memory utilisation refers to the proportion of the total amount of that resource that is available in the DC which is currently allocated to some request. We also observe utilisation metrics relating to each tier of the network, as well as the characteristic relationship between request size and how distributed it is throughout the DC for a particular method. These (baseline or agent). These observations are considered as emergent features (rather than performance-based metrics used to evaluate allocation policies), and are used to try to understand the nature of what each method (the proposed RL-based one in particular) does to achieve the allocation outcomes they do. 
%copy-paste end

\subsection{Baselines}

% The following baselines are those against which the method described in this thesis are compared. One is a trivial random allocator, used only to serve as a presumed lower-limit benchmark but not to suggest any meaningful favourable performance of the method presented. The other methods are taken from literature where network aware resource allocation algorithms have been described. In particular, the NALB and NULB are methods designed for specifically resource-disaggregated data centre allocation. The third baseline, Tetris, while not designed for resource-disaggregated systems specifially, is designed for a more general notion of bandwidth/connectivity aware resource packing in multi-resource data centre environments. As such it is believed to be reasonably compared against the method presented here and the NALB and NULB methods in the resource-disaggregated data centre environment scenario that is addressed. 

% Similar to the process described in section \ref{section:mdp}, each allocation algorithm's allocation choices are successful or failed by the same method. Each time a server is chosen, it is checked if that server can be connected to the servers previously allocated to that request using the k-shortest paths routing procedure. Requests fail when this is not possible.

\subsection*{Tetris}
Tetris \cite{Grandl14} is a multi-resource packing heuristic. It uses the cosine similarity between task requirement and server resource availability vectors to calculate scores upon which packing decisions are made. Network resources are considered to be those which are present at a particular server (e.g. how many free communication channels are available at a particular node), but does not account for less local network resources (e.g. resources across links n-hops away). It also imposes a score penalty on non-local resources in order to encourage locality in its decision making, whereby given an initially chosen server, the score of non-local (not in the same rack) will be slightly decreased. In this way an assumption about network resource efficiency is imposed which suggests that it is better to keep allocations rack local more than not.
\subsection*{NALB}
NALB is a network-aware resource allocation algorithm which uses a bandwidth-weighted breadth-first-search algorithm and a bandwidth-weighted k-shortest paths routing algorithm to find suitable nodes and establish connectivity between them respectively \cite{Zervas18}. The algorithm accounts for CPU, memory, storage (not used in this work), bandwidth and/or latency, where relative weighting between bandwidth and latency is a tunable parameter of the algorithm. The resource environment in the work presented in this thesis is `heterogeneous' with respect to each server (i.e. each server contains various resource types). As such the first server is chosen with a vectorised fitting procedure, similar to the Tetris method described above, instead of the original resource-contention based method designed for resource-homogeneous servers. Otherwise, all implementation features are identical to those presented in the paper introducing this method.
\subsection*{NULB}
NULB works similarly to the NALB method, except that the breadth-first-search algorithm does not use weighting. Weighting is still used for the k-shortest paths procedure \cite{Zervas18}.
\subsection*{Random}
Servers are selected randomly. Used as a lower bound on expected performance.
\definecolor{colorA}{rgb}{0.12156862745098039, 0.4666666666666667, 0.7058823529411765}
\definecolor{colorB}{rgb}{1.0, 0.4980392156862745, 0.054901960784313725}
\definecolor{colorC}{rgb}{0.17254901960784313, 0.6274509803921569, 0.17254901960784313}
\definecolor{colorD}{rgb}{0.8392156862745098, 0.15294117647058825, 0.1568627450980392}
\definecolor{colorE}{rgb}{0.5803921568627451, 0.403921568627451, 0.7411764705882353}

\section{Results}

\subsection{DRL agent allocates more requests overall}

On testing, the agent is observed to consistently outperform every baseline across each topology tested. For each topology, the percentage by which the agent improves over the best performing baseline on that topology is shown in Table \ref{table:comparison}. Most notably, it is seen that the agent thrives in particular when the network has few channels-per-links and/or when oversubscription is high (i.e. when the network is generally resource-constrained), achieving a 19.0\%, 24.4\% and 21.7\% improvement for acceptance, CPU utilisation and memory utilisation respectively. Moreover, the agent is also able to find improvements even in the least resource-constrained environment where even the random baseline is comparable to some of the other baselines. It is also seen that, unsurprisingly, the agent achieves similarly improved resource utilisation for CPU and memory. In this case the same respective improvements are 5.8\%, 2.7\% and 2.7\%. This is a natural emergent outcome of allocating more requests; higher acceptance ratio is equivalent to more requests occupying resources in the DC on average in a given moment in time. As such resource utilisation will also be higher on average.

% Beyond a clear ability to improve over the baselines, this simple view of the performance is suggestive of two key aspects of performance benefits in particular; adaptability and consistency. Adaptability refers to the ability of the RL method to allocate differently in different topologies, which may each have different constraining factors. Consistency refers to the observation that the RL is \textit{always} better, rather than \textit{sometimes} better than the baselines. These aspects will be more clearly indicated by further analysis of the performance of the RL method vs the baselines throughout this section.

\subsection{DRL agent is more consistent than baselines across different DCNs}

\begin{figure*}[h]
    \centering
    % \begin{subfigure}
        \includegraphics[width=0.3\textwidth]{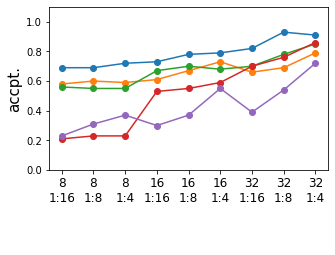}
    % \end{subfigure}
    % \begin{subfigure}
        \includegraphics[width=0.3\textwidth]{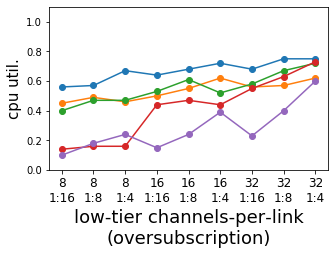}
    % \end{subfigure}
    % \begin{subfigure}
        \includegraphics[width=0.3\textwidth]{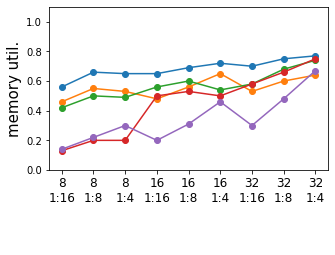}
    % \end{subfigure}
    \caption{\textbf{\textcolor{colorA}{blue}=RL agent, \textcolor{colorB}{orange}=Tetris, \textcolor{colorC}{green}=NALB, \textcolor{colorD}{red}=NULB, \textcolor{colorE}{purple}=random.} Line plots showing the acceptance ratio (top), CPU utilisation (middle) and memory utilisation (bottom) for each method when tested on each topology. Topology labels (x-axis) are defined as $\frac{channels\ per\ tier-1\ links}{oversubscription}$}
    \label{fig:line_resource}
\end{figure*}

The average performance at the end of the test episode for acceptance, CPU utilisation and memory utilisation are shown in Fig. \ref{fig:line_resource}. A key observation from these plots relating to both consistency and flexibility benefits is that while the RL method is always the best performing method on each topology, the baselines are frequently trading places for $2^{nd}$, $3^{rd}$, $4^{th}$ and $5^{th}$.

As previously noted, heuristics are designed on the basis of some specific assumptions about a given system or problem, and are also tested in limited conditions. For example, the Tetris baseline assumes a statically defined emphasis on locality is beneficial, and also asserts that the network and node resources accounted for when scoring a particular server should be only it's local ones (i.e. it's directly attached resources, as opposed to accounting for resources from nodes/links up to k-hops away, for example). This is not so say that heuristics are entirely inflexible; Tetris parameterises how much of a penalty non-local servers should receive, and NALB parameterises the weighting between latency and bandwidth in the routing process for when both features are used. However, the fundamental decision making processes as well as what information is used to make these decisions are for the most part static after the design and testing phase. In this sense a heuristic has some inherent bias in it's behaviour that is derived from the design assumptions and test-performance feature-tuning, and as such do not necessarily have consistent performance benefits over some other heuristic in every circumstance. The tests who's results are shown in Table \ref{table:comparison} and Fig. \ref{fig:line_resource} differ only by the per-link resource quantity and share the fundamental topology. Even so, relatively simple variation is already enough to show the inconsistency of heuristics in this regard. Conversely, the agent learns appropriate policies for each network-resource profile and is able to consistently find better performance across each topology.
%copy-paste start
\begin{table}
\caption{Percentage improvement of the agent pair over the second best performing baseline for that topology across all tested topologies.}
\label{table:comparison}
\begin{tabular}{lllllllllll}
\multicolumn{11}{c}{\textbf{RL agent improvement over best baseline (\%)}} \\
\multicolumn{11}{c}{\textbf{(acceptance, CPU util., memory util.)}} \\
\toprule
\multicolumn{11}{c}{\textbf{Oversubscription}} \\
\midrule
{} & {} &          \multicolumn{3}{c}{1:16} &          \multicolumn{3}{c}{1:8} &          \multicolumn{3}{c}{1:4}        \\
% {} & {} & \multicolumn{3}{c}{acp, cpu, mem} & \multicolumn{3}{c}{acp, cpu, mem} & \multicolumn{3}{c}{acp, cpu, mem}\\
\midrule
 \textbf{Chan.} & 8  & \multicolumn{3}{c}{ 19, 24, 22} & \multicolumn{3}{c}{ 15, 16, 20} & \multicolumn{3}{c}{ 22, 43, 23} \\
 \textbf{per-link} & 16 & \multicolumn{3}{c}{  9, 21, 16}  & \multicolumn{3}{c}{ 11, 12, 15}  & \multicolumn{3}{c}{  8, 16, 11} \\
 \textbf{tier-1} & 32 & \multicolumn{3}{c}{ 17, 17, 21} & \multicolumn{3}{c}{ 19, 12, 10} & \multicolumn{3}{c}{    6, 3, 3} \\
\bottomrule
\end{tabular}
\end{table}

\subsection{DRL agent is more consistent than baselines with respect to request size}

The plots shown in Fig. 8 (Appendix B) show the number of successful allocations of requests per method, where request sizes are grouped relative to the number of total servers worth of resources that their specification rounds up to. This value is an integer between 1 and 8 inclusive. The reward structure is designed to be minimally imposing on the kind of policies the agent can learn, and as such is related only to how many requests it successfully allocates rather than some request-specific information (e.g. the resource requirement magnitude). This was done to attempt to influence the agent to learn policies that are `fair' with respect to any request it encounters. 

% Consider the `shortest remaining processing time' scheduling heuristic. In practise, this heuristic (compared to first-fit, for example) will lead to a higher number of jobs being processed in a given amount of time \cite{parsonson21}. However, the nature of the heuristic is to prioritise short jobs, so in theory a situation can arise where many jobs are completed but no long-running job is ever completed. As such the un-elaborated metric of simply how many jobs were completed is misleading when evaluating the performance of this metric. Since the agent is exposed to a uniformaly distributed set of request sizes, the agent will see less of the extreme sizes (both small and large) during training. As such, it must be considered whether the simple acceptance ratio based reward would therefore incentivise an agent that will prioritise the `middle' size requests at the neglect of smaller/larger ones?

We expect from this design choice that the request should not learn to treat any particular size-range of request more carefully than others. To this end we analyse the number of accepted requests (per topology) aggregated over all test episodes and grouped by request size and explore whether the agent's advantage over the baselines is concentrated in certain kinds of requests, or distributed over all request sizes. We group requests by the minimum number of servers required to fulfill that resource request (1-8 inclusive). What is seen is that the agent's advantage is very consistent across not just topologies but also request size brackets. The only exceptions are for the 3 most resource-constrained topologies, where the agent incurs a small deficit in the largest request size brackets. The worst case is seen on the 8-channel 1:8 oversubscription ratio topology, where deficit of 10 and 2 requests for the requests in the 7-server and 8-server size brackets respectively. This amounts to 1.8\% of the total number of requests received across this test, compared to an overall improvement of 15\%, where all other request-size brackets are improved relative to the baselines.

\subsection{The agent requires less networking resources for the similar allocation performance}

% There are 3 graphs associated with a specified number of channels-per-link at Tier-1 - one for each tested oversubcription ratio. Each of these topologies has the same amount of network resources (channels) at Tier-1 since the the number of racks, servers-per-rack and links-per-server are the same. As such, the networking resources differ in the upper 2 tiers of the network - this is where the different oversubscription ratios emerge. In this way, given the set of topologies with Tier-1 links have $n$ channels-per-link, the 1:8 oversubscribed topology has $2\times$ more networking resources than the 1:16 oversubscribed, and the 1:4 oversubscribed topology has $2.2\times$ more. 

% For the sets of topologies with 8, 16 and 32 channels on links connecting servers to racks respectively, compared to the best performing baseline on the 1:4 oversubscribed topology the agent's performance on the 1:16 topology was 17\% better, equal and 5\% worse. This shows the agent to be able to allocate requests with similar (or better) ability whilst requiring $2.2\times$ less resources, particularly in the cases when network resources are more limited. Similarly, the agent achieves an acceptance ratio on the 1:16 oversubscribed topology with 8 channels-per-link at Tier-1 that is only 1\% less than the best baseline on the 1:16 topology with 32 channels-per-link, requiring $3\times$ less networking resources for approximately the same acceptance.

The agent's performance is also favourable compared against the baselines not just on the same topologies, but also across topologies. In particular, the agent can enable higher acceptance ratio on lower-resource topologies than the best performing baseline on higher-resource topologies. In the most extreme case observed across all experiments, the agent achieves an acceptance on the 8-channel 1:16 oversubscription topology that is only 1\% lower than what the best performing baseline achieves on the 32-channel 1:16 oversubscription topology. In this case, the agent is effectively allowing for the same resource allocation service level to be achieved with $3\times$ less resources.

Practically speaking, this is a very desirable feature. While optical DCN networks can allow various scalability issues to be avoided, the disaggregated resource paradigm that they enable imposes heavy demand on the network. An allocation policy that is able to minimise the amount of networking resources required to maintain some level of resource efficiency can allow for such systems to be built more feasibly. Large network infrastructure requirements (e.g. lots of fibre and switches) is expensive and requires much maintenance and planning \cite{Poutievski22}. Minimising these requirements is highly desirable to limit initial capital, operational and maintenance costs of such systems.

% RL-based method completes in maximum $N$ steps for an $N$-server RDDC, where both the message passing process and inference process can be parallelised. Tetris is also linear with respect to number of nodes and paralellisable, but NALB and NULB are recursive and so have super-linear runtime-scaling with respect to the size of the RDDC topology.

\subsection{Topology scale-up performance}

% Each agent was rolled out on a topology with the same oversubscription properties as the one it was exposed to during training, but with significantly more nodes and over a much longer episode. Specifically, the larger topologies had $\mathcal{O}(10^3)$ nodes and episodes lasted for $\mathcal{O}(10^3$ requests. Otherwise, all oversubscription structures and channels-per-link were the same as the smaller topologies. The rollout is possible due to the architectural choices made for the DRL policy - it is based on a GNN architecture who's NN-model is sized independently to the size of the topology of the graph for which it is generating node embeddings. As such it can be directly applied with no further modification \cite{Hamilton2017}.

Table \ref{table:delta} shows the percentage delta in test performance for the DRL agent when applied to each topology vs it's $\mathcal{O}(10^2)$ scaled up version. In the table $+ve$ indicates that the large-topology score was better and $-ve$ indicates that it was worse. Averaged across each topology, the delta is $5.6\%$, $3.4\%$ and $6.5\%$ for acceptance, CPU utilisation and memory utilisation respectively. The key indication from these results is simply that the agent is clearly able to learn a policy on a small graph that accounts for features of the topology and request distribution that are valid when applied to much larger topologies and over a much longer series of requests.

% In practical context, this is also important should such an approach to allocation be implemented in a future DC. The agent has to learn by means of interacting directly with some kind of allocation environment. In the case that a highly accurate simulation is not available (which is commonly the case for complex systems such as DCNs \cite{Sharma2011,Cortez2017}) this would otherwise require for some real cluster resources to be used for testing. Since the DRL agent begins training randomly initialised it's performance will be very poor (effectively a random allocation policy). In this sense it is of course not feasible to portion some small cluster within the real DCN for training purposes, as this portion would yield extremely poor service to the customers who's requests were allocated to it. However, it is possible to portion a small cluster and mimic it's real commercial operation by using traces from historical request data, which is known to the DCN operator. Nevertheless, doing so is expensive and can certainly not be done at the full DCN scale. As such, in reality, it is quite essential for some learning based policy to be able to learn on a small cluster before being reliably deployed at scale to a full one. The stability of the policy under scale seen here (elaborated in section \ref{section:interpretation}) suggests that this requirement is met, making real-life implementation more feasible.

\begin{table}
\caption{}
\label{table:delta}
\begin{tabular}{lllllllllll}
\multicolumn{11}{c}{\textbf{RL agent performance delta on larger topologies (\%)}} \\
\multicolumn{11}{c}{\textbf{(acceptance, CPU util., memory util.)}} \\

\toprule
\multicolumn{11}{c}{\textbf{Oversubscription}} \\
\midrule
{} & {} &          \multicolumn{3}{c}{1:16} &          \multicolumn{3}{c}{1:8} &          \multicolumn{3}{c}{1:4}        \\
% {} & {} & \multicolumn{3}{c}{accpt, cpu, mem} & \multicolumn{3}{c}{accpt, cpu, mem} & \multicolumn{3}{c}{accpt, cpu, mem}\\
\midrule
 \textbf{Low-tier} & 8.0  & \multicolumn{3}{c}{ 6, 4, 0} & \multicolumn{3}{c}{ 6, -12, -2} & \multicolumn{3}{c}{4, 13, 6} \\
 \textbf{channels} & 16.0 & \multicolumn{3}{c}{  16, 8, 14}  & \multicolumn{3}{c}{ 10, 3, 10}  & \multicolumn{3}{c}{9, 4, 11} \\
 \textbf{per-link} & 32.0 & \multicolumn{3}{c}{ 7, 4, 11} & \multicolumn{3}{c}{ -6, 8, 3} & \multicolumn{3}{c}{-1, -1, 6} \\
\bottomrule
\end{tabular}
\end{table}

% \begin{figure}[h!]
%     \centering
%     \includegraphics[width=\textwidth]{figs/small_top_bar.png}
%     \caption{Histogram of the difference between the number of accepted and rejected requests relative to the number of CPU + memory units requested, shown across results from all tests combined.\newline \textbf{Legend: \textcolor{colorA}{blue}=RL agent, \textcolor{colorB}{orange}=Tetris, \textcolor{colorC}{green}=NALB, \textcolor{colorD}{red}=NULB, \textcolor{colorE}{purple}=random.}}
%     \label{fig:bar}
% \end{figure}

\subsection{Interpreting the policy's allocation strategy}
\label{section:interpretation}

\begin{figure*}[h]
    \centering
    % \begin{subfigure}
        \includegraphics[width=0.3\textwidth]{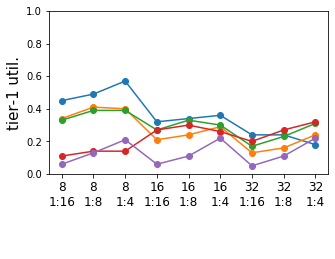}
    % \end{subfigure}
    % \begin{subfigure}
        \includegraphics[width=0.3\textwidth]{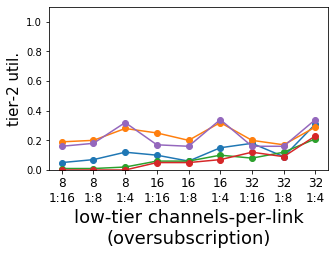}
    % \end{subfigure}
    % \begin{subfigure}
        \includegraphics[width=0.3\textwidth]{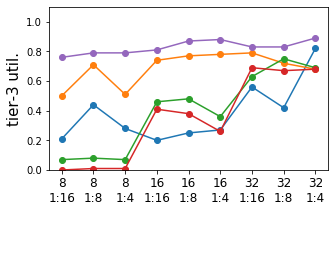}
    % \end{subfigure}
    \caption{\textbf{\textcolor{colorA}{blue}=RL agent, \textcolor{colorB}{orange}=Tetris, \textcolor{colorC}{green}=NALB, \textcolor{colorD}{red}=NULB, \textcolor{colorE}{purple}=random.} Line plots showing the utilisation of tier-1 (top), tier-2 (middle) and tier-3 (bottom) networking resources for each method when tested on each topology. Topology labels (x-axis) are defined as $\frac{channels\ per\ tier-1\ links}{oversubscription}$}  
    \label{fig:line_network}
\end{figure*}

Here is presented a discussion, led by visual analysis, on the nature of the allocation policy that is learnt by the DRL agent. Since it is unclear how to directly extract policy principles from neural networks directly, numerical and visual analysis through experimental probing is required to infer as best as possible what the policy is doing. In effect, this section attempts to describe the learnt allocation policy as if it were a heuristic.

\subsubsection{DRL agent uses network when it is available}
\label{section:interpretation_network}

Figure \ref{fig:line_network} shows the utilisation of each tier's network resources per-topology-per-method. Looking at the agent's results, it is seen that when the network resources are very limited (highly oversubscribed and few channels-per-link), the method concentrates allocations within racks, having higher utilisation for tier-1 (intra-rack) and comparatively very low utilisation for the higher tiers. As the agent moves towards topologies with lower oversubscription /more channels-per-link, the tier-2 and tier-3 resources are more highly utilised and the tier-1 resources less. This indicates that the agent learns a policy that exploits network resources when they are available, but allocates more rack-locally when they are not. Moreover, the biggest increase in utilisation occus at tier-3, which increases from $\approx 0.2$ to $\approx 0.8$ between the most and least network-resource constrained topologies. This indicates that in particular the agent learns to exploit the most non-local allocation possible (inter-cluster) when network resources are available to do so.

Similarly, Figure \ref{fig:server_distributioin} provides a more aggregated but intuitive representation of how the policy exploits the network in various oversubscription scenarios. Each distribution shows the combined results of allocation outcomes for all topologies of a particular oversubscription ratio (per-method). The x-axis refers to how distributed the servers allocated to a request were, and the y-axis indicates how commonly that distribution was used by an allocation method. Distribution outcomes (intra-server, intra-rack, intra-cluster and inter-cluster) are presented continuously rather than discretely, since mixtures are generally possible (i.e. it might be the case that 90\% of the servers allocated to a request are in the same rack and the other 10\% are in a neighbouring rack meaning that the request is mostly intra-rack and partially intra-cluster). Observing how the distributions change moving from oversubscription of 1:16 through to 1:8 and 1:4, a similar conclusion as described above can be drawn. In the 1:16 oversubscribed topology, the agent is almost entirely intra-rack. As the network opens up more in the 1:8 oversubscription case it can be seen that more requests are distributed intra-cluster (across neighbouring racks) but inter-cluster allocations are insignificant. Finally in the most network-resourced topology with 1:4 oversubscription, the agent reduces it's dependency on intra-rack allocations significantly, allocating nearly as many intra-cluster and inter-cluster as it does intra-rack (though intra-rack still remains the dominant allocation outcome due to racks having the most network resources per connected server. By comparison, the distributions for each heuristic are notably much more static when self-compared across oversubscription ratios, since it is the exact same policy deployed in each case. While Tetris changes slightly between 1:8 and 1:4 oversubscription to exploit intra-cluster with a slightly high proportion, the NALB and NULB maintain almost exactly the same shape indicating that their network-usage is independent of what is actually available.

\subsubsection{The agent distributes requests differently based on their resource requirements}

\begin{figure}[h]
    \centering
    % \begin{subfigure}
        \includegraphics[width=0.15\textwidth]{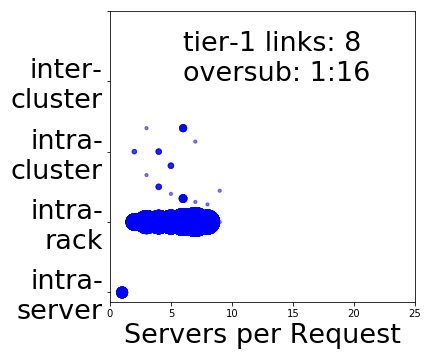}
    % \end{subfigure}
    % \begin{subfigure}
        \includegraphics[width=0.15\textwidth]{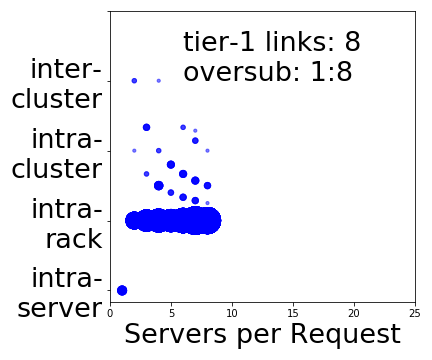}
    % \end{subfigure}
    % \begin{subfigure}
        \includegraphics[width=0.15\textwidth]{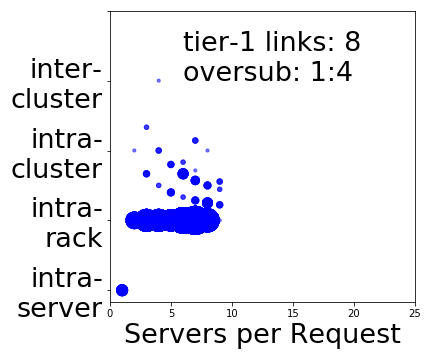}
    % \end{subfigure}
    % \begin{subfigure}
        \includegraphics[width=0.15\textwidth]{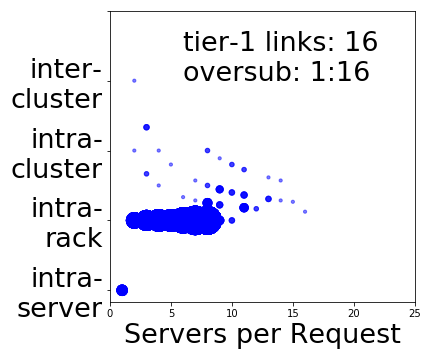}
    % \end{subfigure}
    % \begin{subfigure}
        \includegraphics[width=0.15\textwidth]{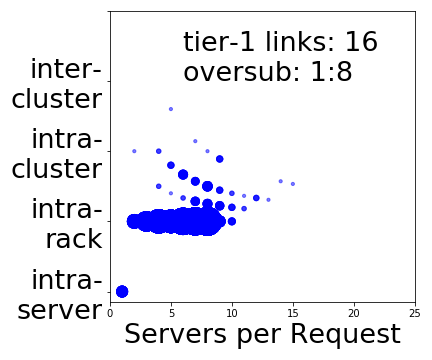}
    % \end{subfigure}
    % \begin{subfigure}
        \includegraphics[width=0.15\textwidth]{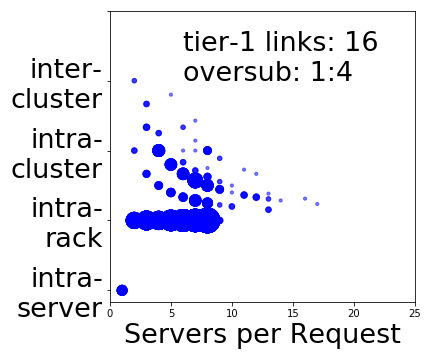}
    % \end{subfigure}
    % \begin{subfigure}
        \includegraphics[width=0.15\textwidth]{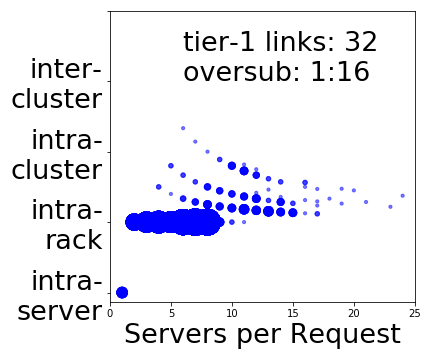}
    % \end{subfigure}
    % \begin{subfigure}
        \includegraphics[width=0.15\textwidth]{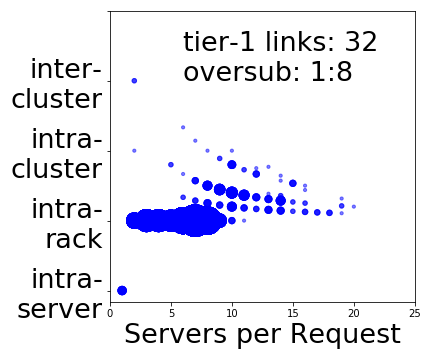}
    % \end{subfigure}
    % \begin{subfigure}
        \includegraphics[width=0.15\textwidth]{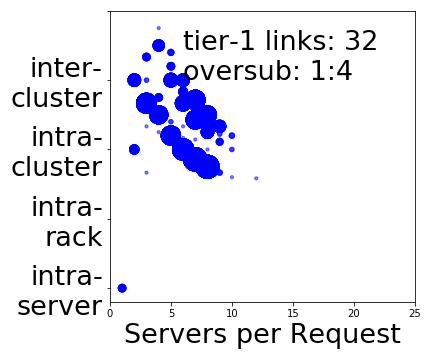}
    % \end{subfigure}
    \caption{Visualising the deep reinforcement learning agent's policy with respect to relationship between how many servers were allocated to a request, and how distributed those servers were for that request. Blue dots represent server-distribution pairs and their size represents how many requests were served at this combination. Shown for each training topology.}
    \label{fig:policy}
\end{figure}

Following the conclusions discussed in section \ref{section:interpretation_network}, further questions can be asked about how the agent learns to exploit network resource availability in relation to the specific request. Specifically, what relationship does the agent learn between request size and how distributed that request should to maximise the possibility of finding acceptable allocations over the long term?

\begin{figure*}[h]
    \centering
    % \begin{subfigure}
        \includegraphics[width=0.3\textwidth]{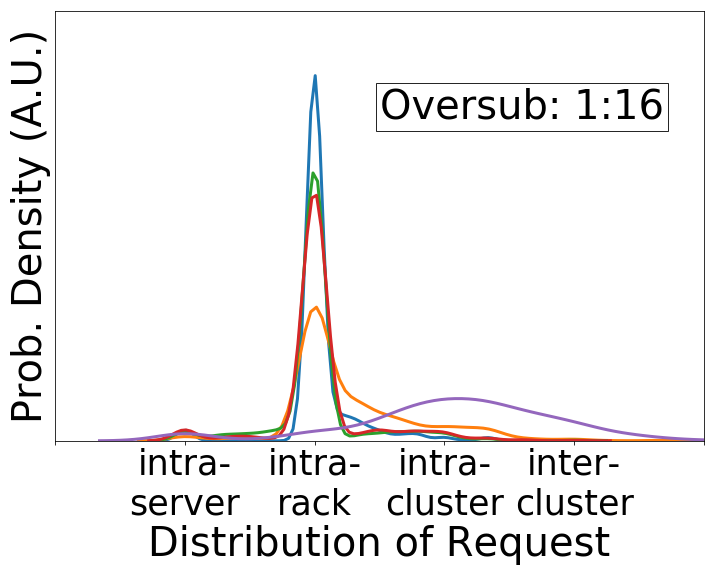}
    % \end{subfigure}
    % \begin{subfigure}
        \includegraphics[width=0.3\textwidth]{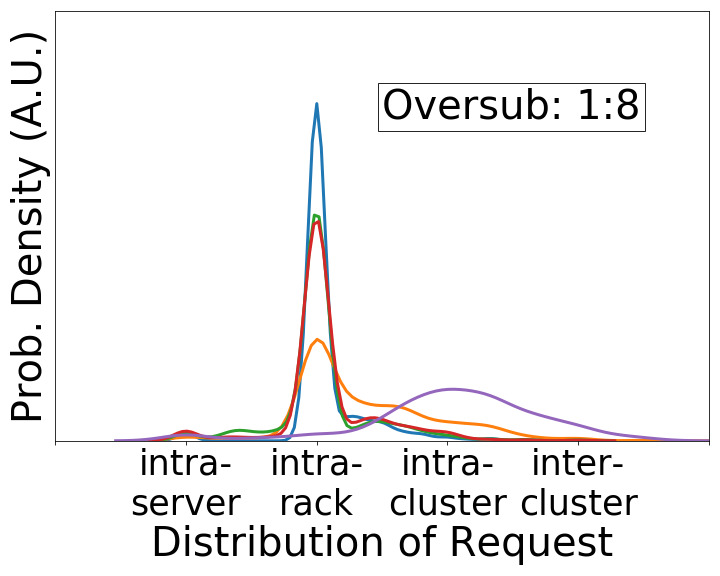}
    % \end{subfigure}
    % \begin{subfigure}
        \includegraphics[width=0.3\textwidth]{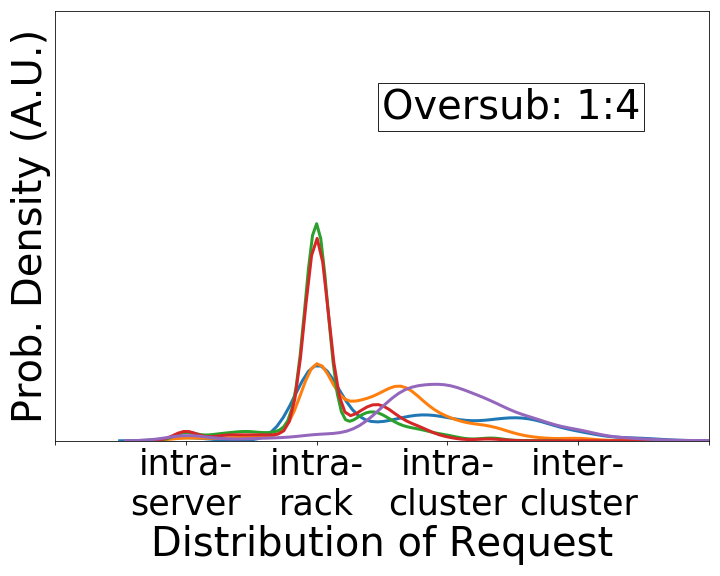}
    % \end{subfigure}
    \caption{\textbf{\textcolor{colorA}{blue}=RL agent, \textcolor{colorB}{orange}=Tetris, \textcolor{colorC}{green}=NALB, \textcolor{colorD}{red}=NULB, \textcolor{colorE}{purple}=random.} Line plots showing the acceptance ratio (top), CPU utilisation (middle) and memory utilisation (bottom) for each method when tested on each topology. Topology labels (x-axis) are defined as $\frac{channels\ per\ tier-1\ links}{oversubscription}$}
    \label{fig:server_distributioin}
\end{figure*}

\begin{figure}[h]
    \centering
    % \begin{subfigure}
        \includegraphics[width=0.15\textwidth]{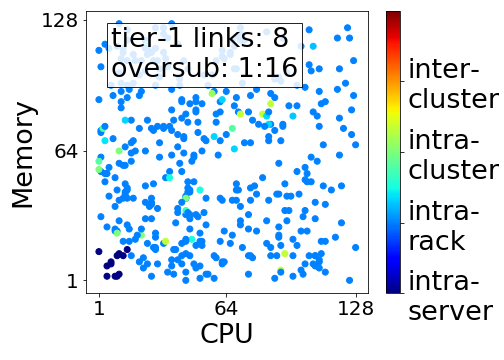}
    % \end{subfigure}
    % \begin{subfigure}
        \includegraphics[width=0.15\textwidth]{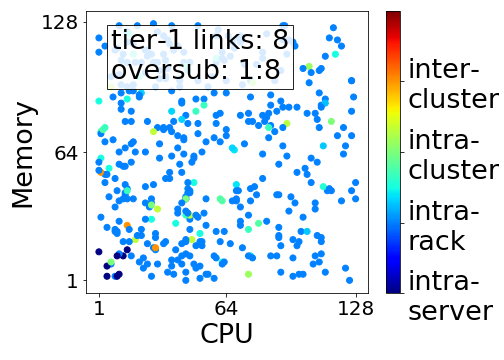}
    % \end{subfigure}
    % \begin{subfigure}
        \includegraphics[width=0.15\textwidth]{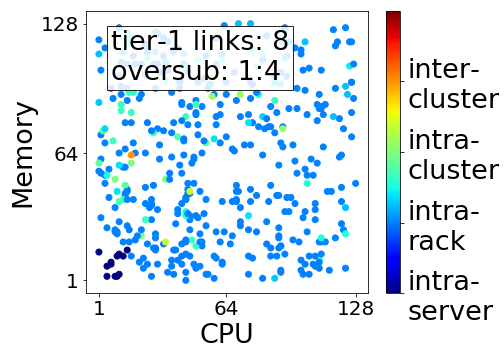}
    % \end{subfigure}
    % \begin{subfigure}
        \includegraphics[width=0.15\textwidth]{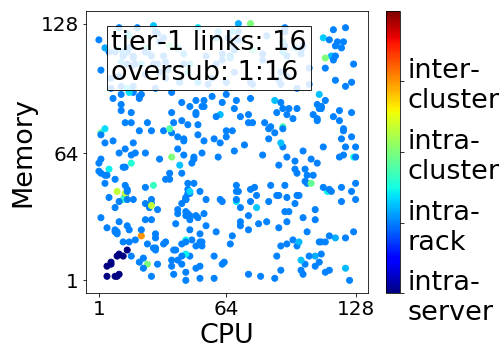}
    % \end{subfigure}
    % \begin{subfigure}
        \includegraphics[width=0.15\textwidth]{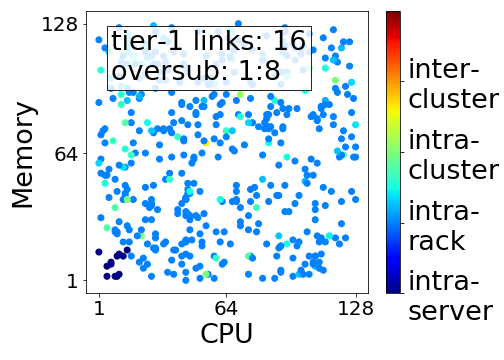}
    % \end{subfigure}
    % \begin{subfigure}
        \includegraphics[width=0.15\textwidth]{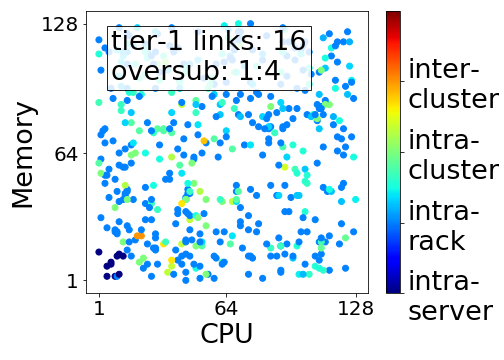}
    % \end{subfigure}
    % \begin{subfigure}
        \includegraphics[width=0.15\textwidth]{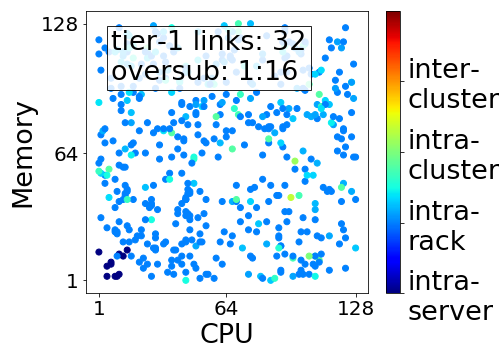}
    % \end{subfigure}
    % \begin{subfigure}
        \includegraphics[width=0.15\textwidth]{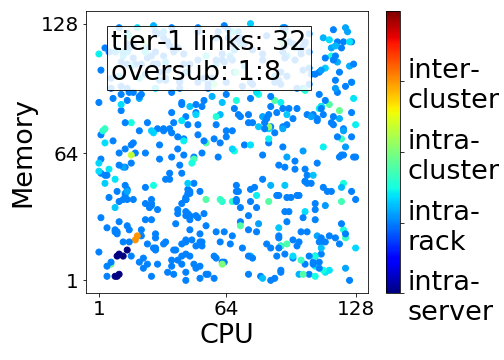}
    % \end{subfigure}
    % \begin{subfigure}
        \includegraphics[width=0.15\textwidth]{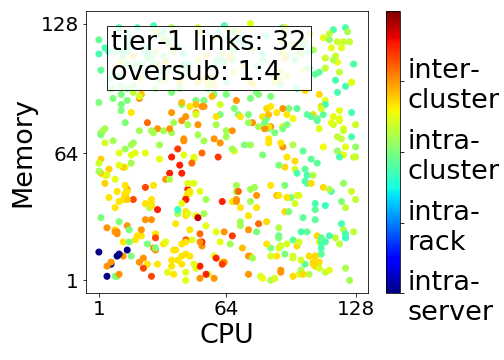}
    % \end{subfigure}
    \caption{Visualising the deep reinforcement learning agent's policy with respect to the relationship between requested CPU units, requested memory units and how distributed the allocated request at that size was. Distribution is represented by colour, where blue to red corresponds to more less to more distributed. Shown for each training topology.}
    \label{fig:resource_2d}
\end{figure}

Figure \ref{fig:policy} shows (for each oversubscription/channels-per-link topology combination) a relationship between how many servers were allocated to a request (x axis) and how distributed those servers were (y axis). In the figures, the size of a blue circle relates to the number of requests which were allocated that number of servers and distributed at that amount. Figure \ref{fig:resource_2d} shows a 2D colour-scaled plot where the x and y axes refer to the number of CPU and memory units requested and the colour of a point indicates how distributed the request with those requirements was distributed. These two plots allow us to consider a relationship between request size, how many servers across which that request was allocated and how those servers were distributed. Note that down-column comparison looks at increasing the channels-per-link quantity, and comparing sub-figures down-row compares decreasing the oversubscription ratio (less oversubscription).

Figure \ref{fig:policy} shows that while there is a preference for intra-rack allocation for on all topologies, a slightly increased preference for greater distribution is seen as the amount of networking resources increases. The most highly distributed requests are also generally the smaller ones using few servers rather than the larger requests with many servers. 
Fig \ref{fig:resource_2d} shows a trend of generally increasing distribution as network resources increase. This trend is more  common among the smaller requests who's likelihood of being more distributed increases most notably with increasing network resources. On the contrary, the largest requests tend to remain predominantly rack-local, even as total networking resources increase. Finally, there is also a general preference for rack-locality since the overall majority of requests (observed over all size brackets) are rack local.

This behaviour can be summarised as 1. rack-local allocations are generally preferred; 2. smaller requests have a higher likelihood of being more highly distributed; 3. larger requests have a higher likelihood of being less distributed and kept rack-local.

Higher tiers of the network are accessible by any server in the DCN. The higher the tier, the more servers are likely to use it's resources. Conversely, lower tiers of the network are likely to be accessed by fewer servers and tier-1 links will only be accessed by their directly attached servers, or a server who wishes to communicate with that server. Larger requests require more servers, and therefore more network resources to interconnect them, whereas smaller requests require fewer network resources correspondingly. What is evident from the policy visualisation and analysis is that the agent learns to tactically use network resources in a way that prevents large requests from congesting the the higher tiers of the network for other requests that arrive in the future. Smaller requests are allocated in a way such that lower-oversubscription (less contended for) regions of the network are kept free for the more demanding larger requests, and smaller requests tend to be more distributed without using excessive higher tier network resources and prohibiting future allocations.

\section{Conclusion}

This paper shows that deep reinforcement learning with graph neural network based policy architectures can be used to learn effective network-aware resource allocation policies end-to-end. When trained and tested across 9 data centre topologies with different network-resource quantity and oversubscription, the presented method achieves up to a 19\%, 24\% and 22\% improvement for acceptance ratio, CPU utilisation and memory utilisation respectively against a number of baseline heuristics for network-aware resource allocation. Improvements are most pronounced when the network resources are most limited. The method also achieves the same performance as the best heuristic whilst requiring $3\times$ less network resources to do so. Additionally, the policy is highly scalable and the policy architecture topology agnostic. When trained on topologies with $\mathcal{O}(10^1)$ servers, policy performance is highly consistent when deployed on topologies with the same oversubscription properties but $\mathcal{O}(10^3)$ more servers with no re-training or architectural adjustments required.

Avenues of future work include training a single agent for a multiple of topologies/network-types; increasing the scale of test topologies beyond the $\mathcal{O}(10^3)$ shown here; handling a wider variety of request types (e.g. different requests requiring different connectivity patterns to all-to-all); greater variety/time-dependent request distributions to be handled during allocation rather than a single static one and more restrictive requests where latency requirements are stated explicitly, not implicitly satisfied by the network and must be accounted for in allocation strategies.

\section*{Acknowledgements}

This work was supported under the Engineering and Physical Sciences Research Council (EP/R041792/1 and EP/L015455/1), the Industrial Cooperative Awards in Science and Technology (EP/R513143/1), the OptoCloud (EP/T026081/1), and the TRANSNET (EP/R035342/1) grants.

\section*{Disclosures}

The authors do not maintain any conflicts of interest related to this project.

% Bibliography
\bibliography{sample}
\appendix

\section{Visualisation of the Training Topology}

\begin{figure}[h]
    \centering
    \includegraphics[width=0.5\textwidth]{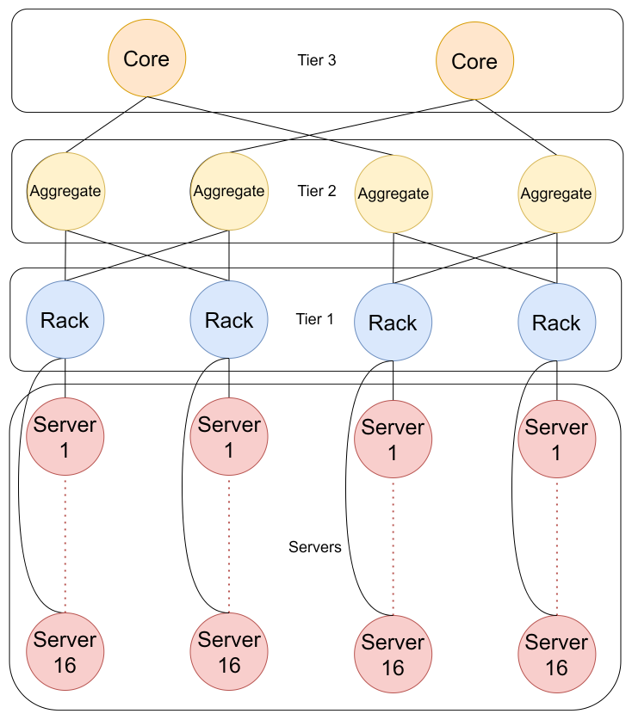}
    \caption{Simple illustration of the data centre network used for training and testing in this work, where the rack-grouping of servers and tier'd structure of the network is labelled explicitly.}
    \label{fig:topology}
\end{figure}

Fig. \ref{fig:topology} shows a simple visualisation of the 3-tier DCN topology used for training. Racks consist of groups of 16 servers, and clusters consist of groups of 2 racks. There are 2 clusters total in this training topology. 

\label{appdx:topology}

\section{Bar Plots for Accepted Requests}
\label{appdx:bar}

\begin{figure}[h]
    \centering
        \includegraphics[width=0.15\textwidth]{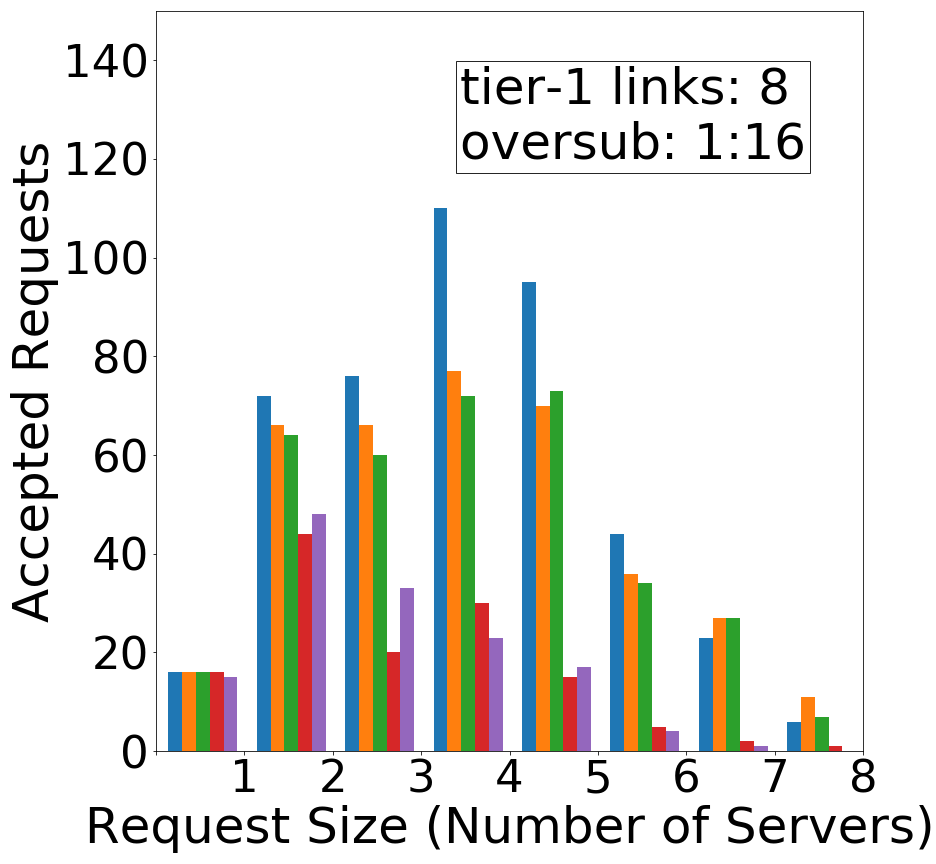}
        \includegraphics[width=0.15\textwidth]{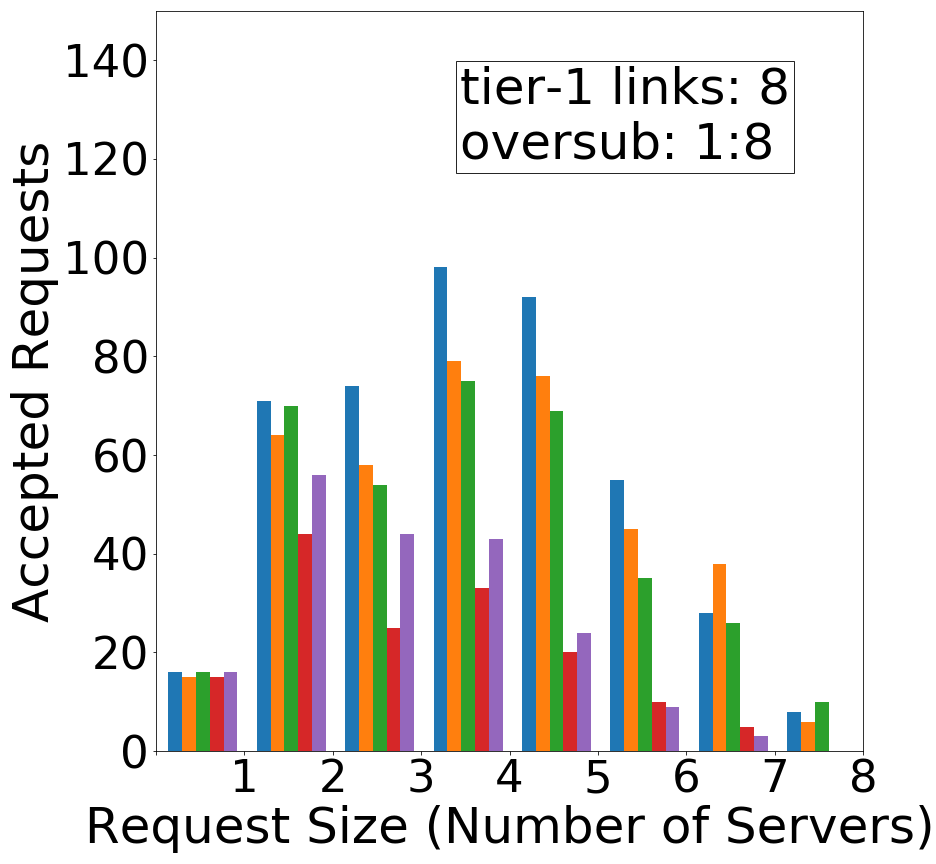}
        \includegraphics[width=0.15\textwidth]{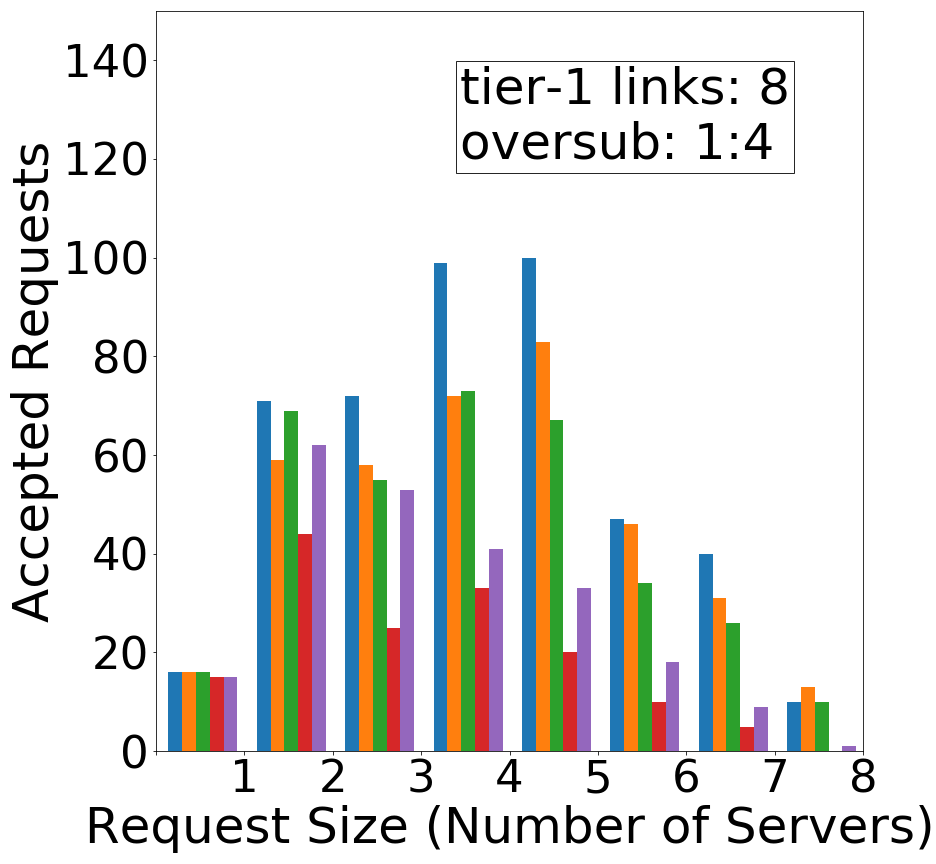}
        \includegraphics[width=0.15\textwidth]{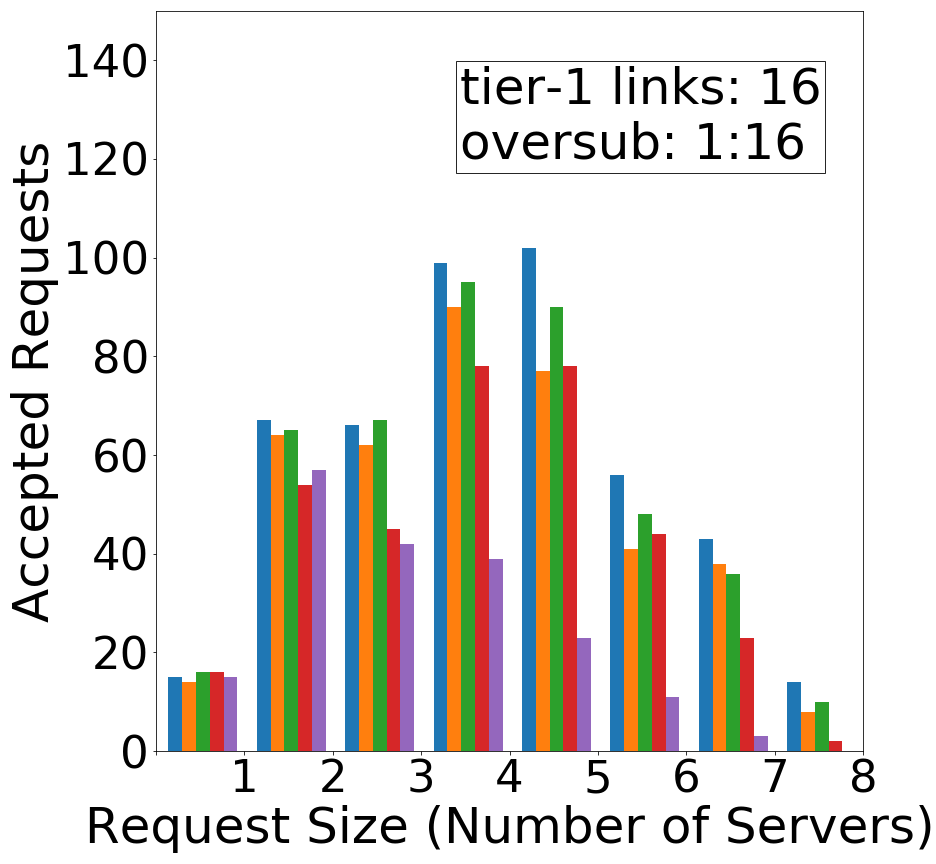}
        \includegraphics[width=0.15\textwidth]{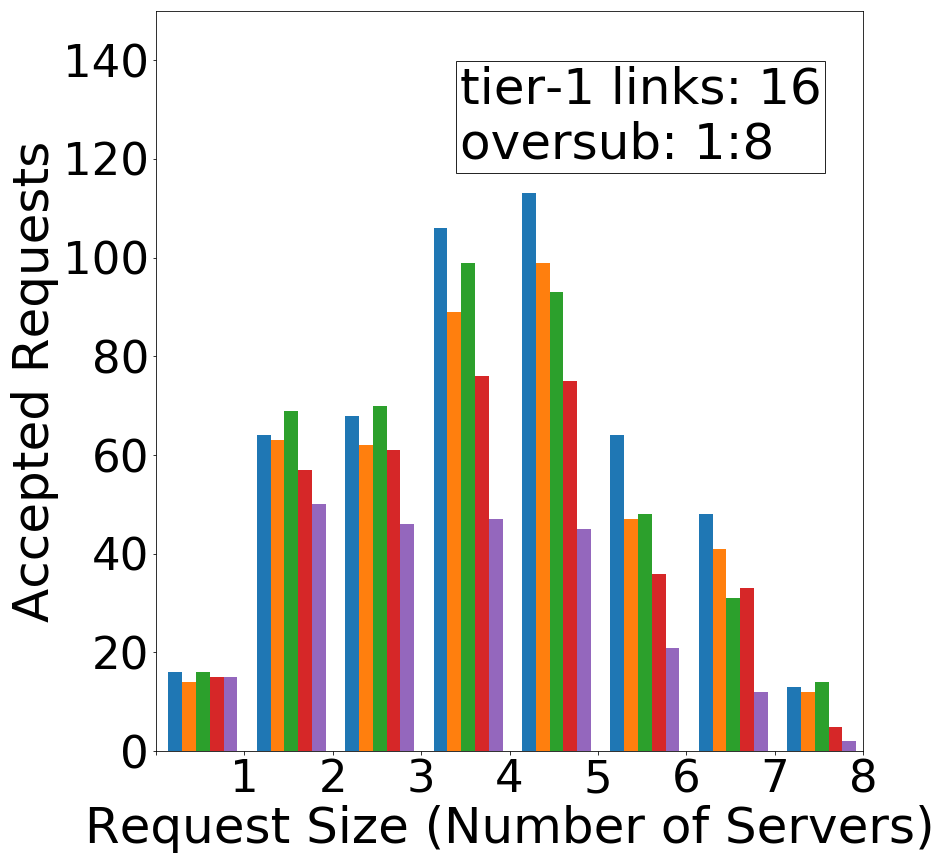}
        \includegraphics[width=0.15\textwidth]{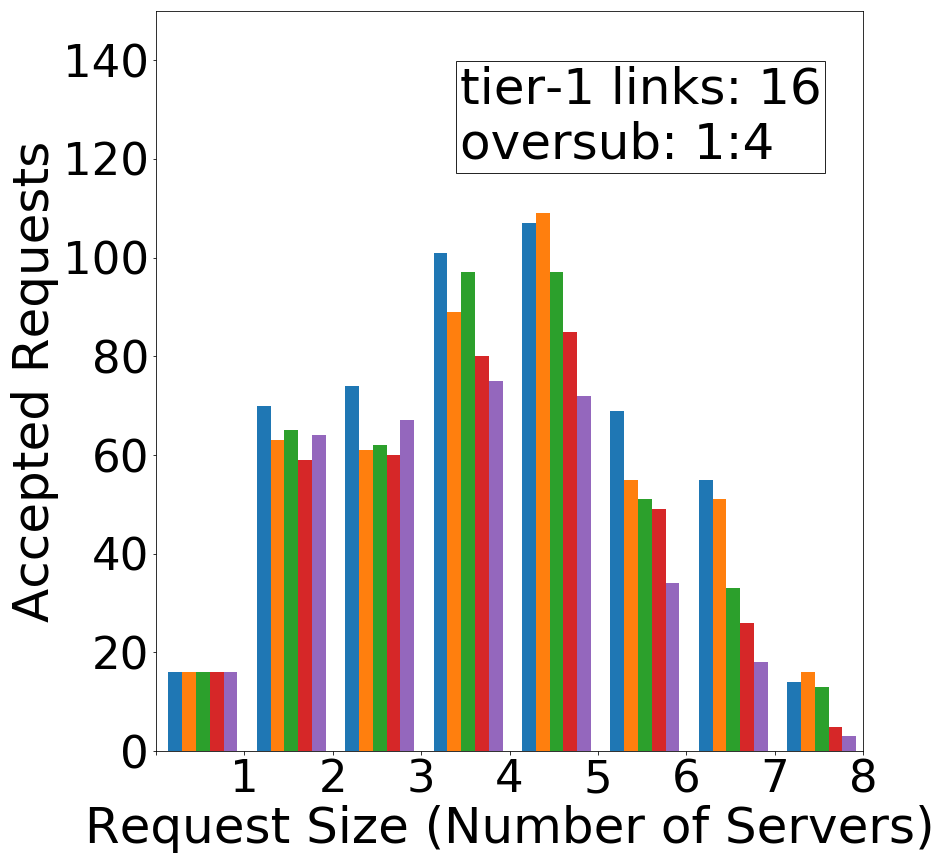}
        \includegraphics[width=0.15\textwidth]{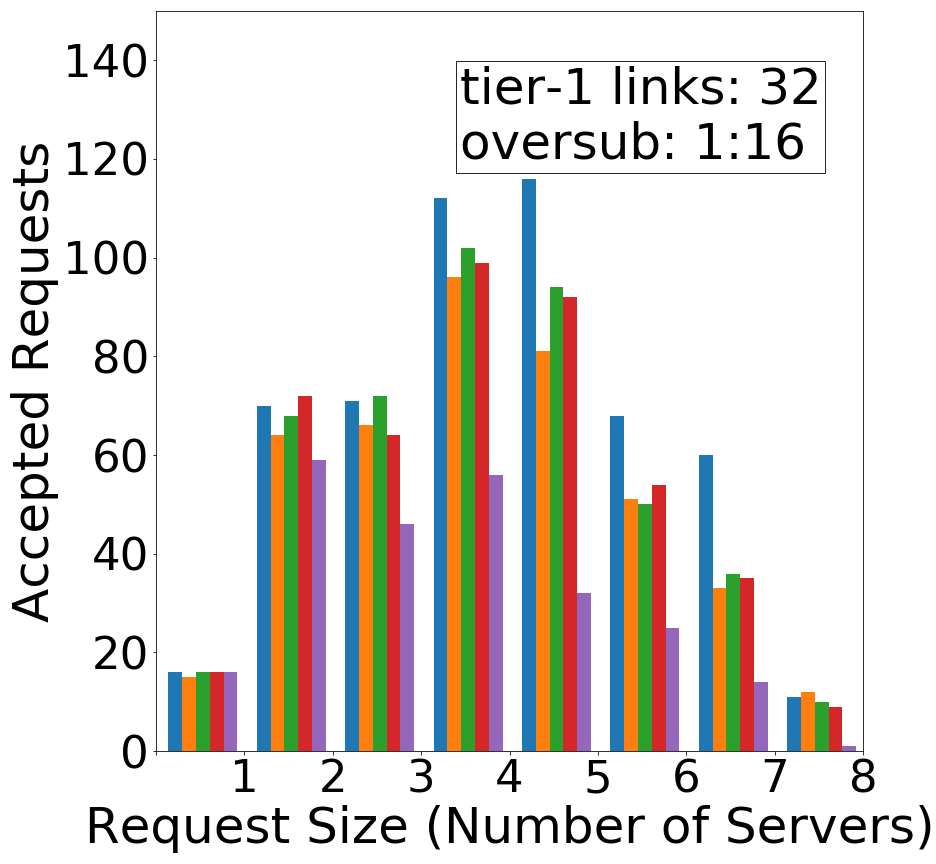}
        \includegraphics[width=0.15\textwidth]{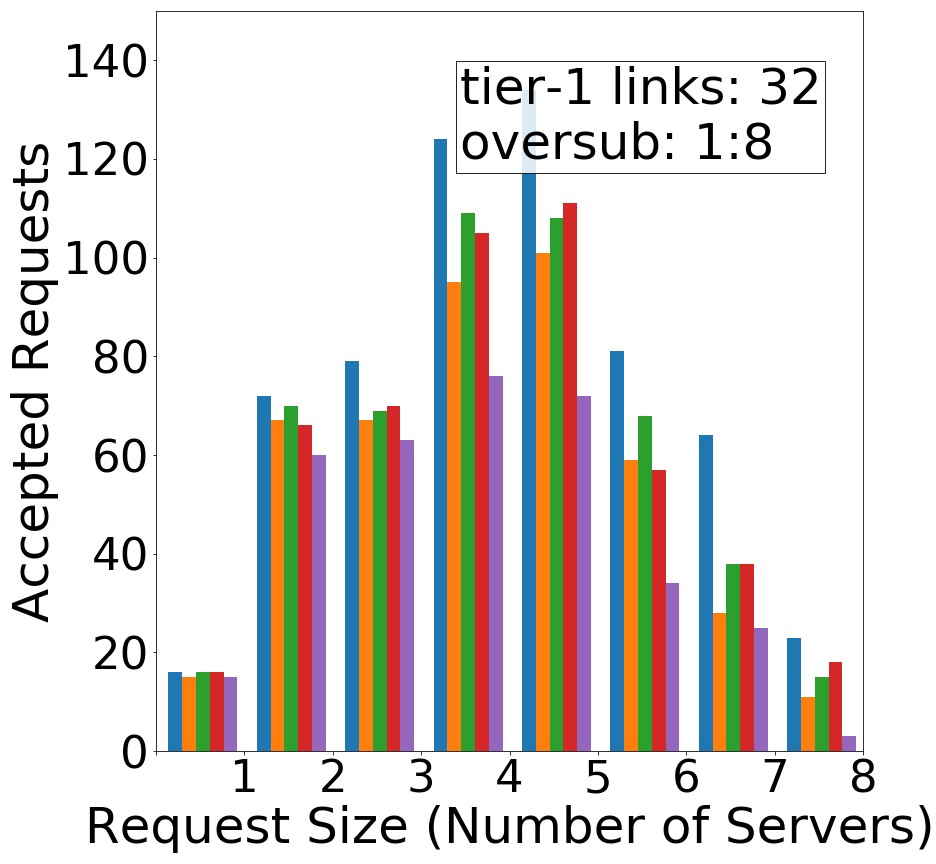}
        \includegraphics[width=0.15\textwidth]{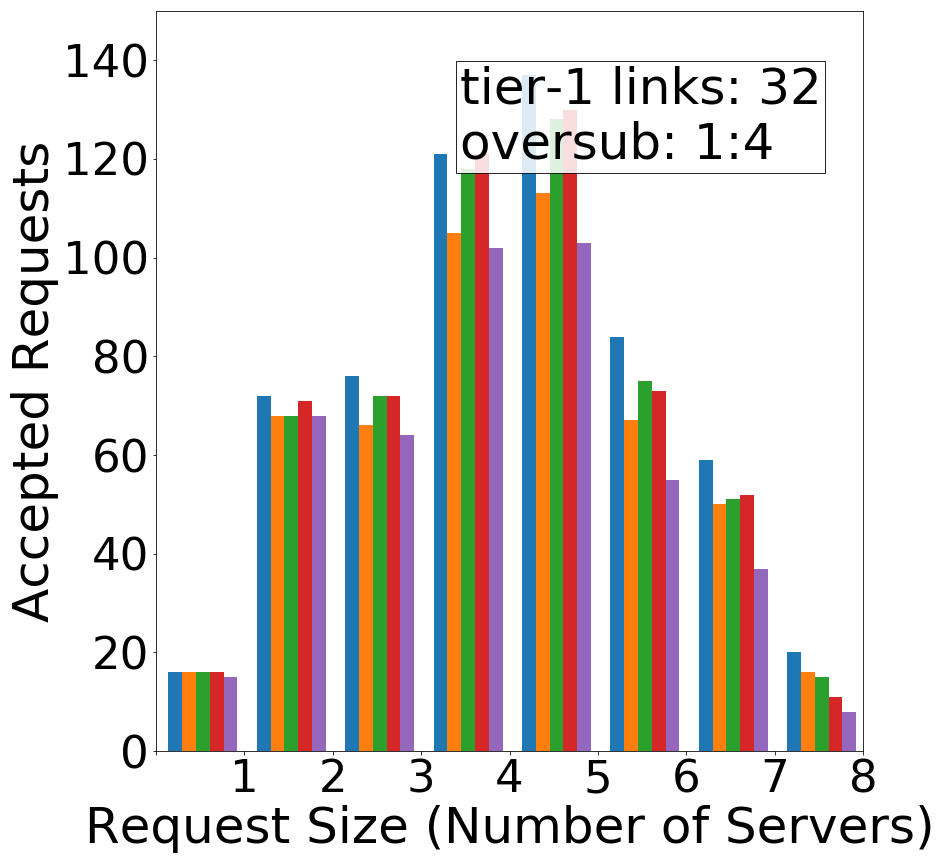}
    \caption{\textbf{\textcolor{colorA}{blue}=RL agent, \textcolor{colorB}{orange}=Tetris, \textcolor{colorC}{green}=NALB, \textcolor{colorD}{red}=NULB, \textcolor{colorE}{purple}=random.} \\ Histograms showing how many requests were successfully allocated by each method, grouped by how many servers worth of total resources were requested (i.e. what is the minimum number of servers that could theoretically fulfil this request). Text on the sub-figures refers to which topology the results are for with respect to it's networking resources and oversubscription (as in the rest of the paper).}
    \label{fig:success_bar_32}
\end{figure}

The plots in Fig. \ref{fig:success_bar_32} show the topology-specific results for allocation success with respect to allocation size (relative to the minimum number of servers required to allocate that request).

\end{document}